\documentclass[twocolumn,showpacs,amsmath,amssymb,superscriptaddress,nofootinbib]{revtex4}

\usepackage{graphicx}
\usepackage{epsfig}
\usepackage{dcolumn}
\usepackage{bm}

\newcommand{\beq}{\begin{equation}}
\newcommand{\eeq}{\end{equation}}
\newcommand{\bea}{\begin{eqnarray}}
\newcommand{\eea}{\end{eqnarray}}

\begin{document}
\title{To what extent does the self-consistent mean-field exist?}
\author{Lu Guo}
\affiliation {Institut f{\"u}r Theoretische Physik, J. W. Goethe-Universit{\"a}t, D-60438 Frankfurt, Germany}
\affiliation {Department of Mathematical Science, Ibaraki University, Mito 310-8512, Ibaraki, Japan}
\author{Fumihiko Sakata}
\affiliation { Institute of Applied Beam Science, Graduate School of Science and Engineering, Ibaraki University, Mito 310-8512, Ibaraki, Japan}
\author{En-guang Zhao}
\affiliation {Institute of Theoretical Physics, Chinese Academy of Sciences, Beijing 100080, China}
\author{J. A. Maruhn}
\affiliation {Institut f{\"u}r Theoretische Physik, J. W. Goethe-Universit{\"a}t, D-60438 Frankfurt, Germany}
\date{\today}

\begin{abstract}
A non-convergent difficulty near level-repulsive region is discussed within the self-consistent mean-field theory.  It is shown by numerical and analytic studies that the mean-field is not realized in the many-fermion system when quantum fluctuations coming from two-body residual interaction and quadrupole deformation are larger than an energy difference between two avoided crossing orbits. An analytic condition indicating a limitation of the mean-field concept is derived for the first time.
\end{abstract}

\pacs{21.60.Jz, 21.10.Pc, 21.30.Fe}

\maketitle

The Hartree-Fock mean-field theory has been greatly successful in
describing many-fermion systems. Level repulsion is
a universal phenomenon appearing in the mean-field description of
molecular, atomic, biological and nuclear systems
\cite{JMA03,RGF03,FI01,CD01}. In the development of nuclear
structure physics, there have been many discussions on the
applicability of cranked mean-field theory near level-repulsive
region \cite{Hama76, Stru77, Dob00}. An argument on how to
remove certain spurious interaction and to construct diabatic orbits
seems to be reasonable in cranked mean-field  \cite{Dob00,
RB89}, because two crossing orbits interact not at a given angular
momentum but at a given angular frequency. However the above
argument may not be simply extended to the deformation constrained
mean-field theory and the interaction between two potential energy
surfaces (PESs) might not be necessarily regarded to be spurious
since the Hamiltonian and deformation operators do not commute with each
other. Although the non-convergent difficulty in the mean-field
calculation was realized numerically \cite{Dob00}, to our knowledge,
there has been no discussion by deriving analytic expressions to
clearly demonstrate why and how the competition between one-body
potential and two-body residual interaction plays a decisive role in
breaking the concept of mean-field near level-repulsive region. For
this purpose, this Letter will give an analytic condition to
indicate the applicability of mean field near level replusion.

\begin{figure}
\epsfxsize=8.3cm \centerline{\epsffile{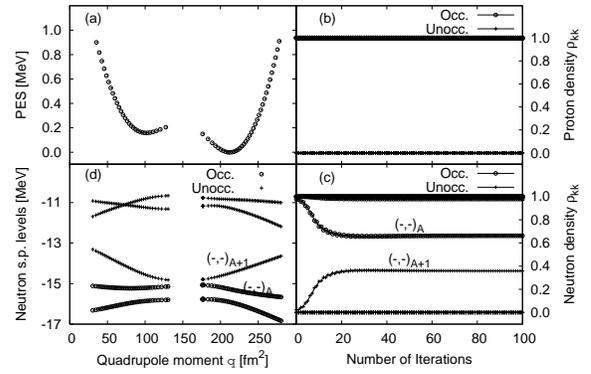}}
\caption{\label{prop} Ground state PES of nucleus $^{66}{\textrm
{Se}}$ (a); diagonal components of proton (b) and neutron (c) densities
as a function of iterative number for a non-convergent case; neutron
s.p. energies near Fermi surface (d).}
\end{figure}

The quadratic constrained Hartree-Fock (CHF) equation with Gogny D1S interaction \cite{Gog4}
has been solved by using three dimensional harmonic oscillator basis
\cite{Guo2}. The triaxial oscillator parameters are determined by
minimizing the energy of HF solution. In the calculation of PES, the
same oscillator parameters as those of the ground state are used to
trace an evolution of the ground state configuration, and to make
the single-particle (s.p.) level-crossing dynamics transparent.
Figure~\ref{prop}(a) shows that there appears a missing region in
the PES of nucleus $^{66}{\textrm {Se}}$. At an edge of the PES
($\langle\hat Q\rangle=q \approx 178$ ${\textrm{fm}}^2$), the CHF
calculation meets a difficulty of non-convergence even one decreases
the deformation quite a little, and no matter how much effort one
makes to get convergence. After the missing region, a
continuously-connected PES passing through another local minimum ($q
\approx 100$ $\mbox{fm}^2$) is obtained. In order to understand what prevents
the CHF calculation from convergence, the diagonal components of
proton (b) and neutron (c) densities as a function of iterative number 
are depicted for a non-convergent case
with $\mu=150$ $\mbox{fm}^2$. Here, $\mu$ is a control parameter of
quadratic constraint given by $-w(\mu-\langle\hat Q \rangle)\hat Q$,
and is used to assign deformation in the missing region instead of
$q$. A representation where one-body CHF hamiltonian $h^{(n)}$ is diagonal is used with $n$ being iterative number.
From Fig.\ref{prop}(b), one may observe that the
expectation values of proton density converge to 0 for unoccupied
orbits and 1 for occupied orbits. For the case of neutron density (c),
there appears a similar situation for most single-hole and
single-particle states, except for two specific orbits labeled as
$(-,-)_A$ and $(-,-)_{A+1}$, where $(-,-)$ denotes parity and signature quantum numbers 
and $A$ the number of occupied orbits. The two specific orbits are interacting and lying just below
and above the Fermi surface as shown in Fig.~\ref{prop}(d). One may
expect that the two specific orbits play a dominant role in preventing the CHF calculation
from convergence.

In what follows, we will focus on how the non-convergent difficulty appears as a result of the microscopic dynamics. Suppose there exists a convergent CHF state at $q=q_0$. We will discuss whether the CHF calculation with slightly different deformation at $q=q_0-\Delta q$  goes to convergence or not. When one takes the CHF state at $q=q_0$ as a starting trial function to proceed an iteration at $q=q_0-\Delta q$, the matrix elements of the CHF Hamiltonian in the first iteration between two specific orbits in a {\it $q_0$-representation} of using the self-consistent s.p. states $\{\epsilon_k(q_0), \varphi_k(q_0)\}$ at $q_0$ are written as (for detailed analytic derivation, see Ref. \cite{Guo3})
\bea
h_{A,A+1}^{(1)}(q_0-\Delta q) &\equiv& \sum_{\alpha\beta}\varphi^{\dagger}_{A\alpha}(q_0) h^{(1)}_{\alpha\beta}\varphi_{\beta A+1}(q_0) \nonumber \\
&=& w\Delta\mu Q_{A,A+1} ,\nonumber \\
h_{A,A}^{(1)}(q_0-\Delta q) &\equiv& \sum_{\alpha\beta}\varphi^{\dagger}_{A\alpha}(q_0)h^{(1)}_{\alpha\beta}\varphi_{\beta A}(q_0) \nonumber \\  &=& \epsilon_A(q_0) + w\Delta\mu Q_{A,A} ,\nonumber \\
h_{A+1,A+1}^{(1)}(q_0-\Delta q) &\equiv& \sum_{\alpha\beta}\varphi^{\dagger}_{A+1\alpha}(q_0)h^{(1)}_{\alpha\beta}\varphi_{\beta A+1}(q_0) \nonumber \\ &=& \epsilon_{A+1}(q_0) +  w\Delta\mu Q_{A+1,A+1} ,
\label{h0d2}
\eea
where $\alpha, \beta,\cdots$ refer to a set of s.p. states used for numerical basis.
In the $q_0$-representation, the matrix elements of deformation operator in Eq.(\ref{h0d2}) are expressed as
\beq
Q_{i,j}\equiv \sum_{\alpha\beta}\varphi^{\dagger}_{i\alpha}(q_0)Q_{\alpha\beta}\varphi_{\beta j}(q_0).
\label{qof}
\eeq
The $2\times 2$ truncated Hamiltonian of $h^{(1)}(q_0-\Delta q)$ in the first iteration is expressed as
\beq
\left( \begin{array}{cc} h_{A,A}^{(1)}(q_0-\Delta q) & h_{A,A+1}^{(1)} (q_0-\Delta q) \\ h_{A+1,A}^{(1)}(q_0-\Delta q) & h_{A+1,A+1}^{(1)}(q_0-\Delta q) \end{array} \right).
\label{hfham}
\eeq

Making the following analytic discussion simple, we exploit such an approximate expression that a contribution to the mean-field from the $(A-1)$ number of hole-states is independent of small deformation change $\Delta\mu$, and of number of iterations, which is well realized in our numerical calculation discussed above. That is
\beq
\sum_{i=1}^{A-1}\varphi_{\alpha i}^{(n)}\varphi_{i\beta}^{\dagger(n)}\approx \sum_{i=1}^{A-1}\varphi_{\alpha i}(q_0)\varphi_{i\beta}^{\dagger}(q_0),\quad n=1,2\ldots .
\label{appeq}
\eeq
Under the above assumption,  one may explore the non-convergent dynamics governing the CHF calculation in terms of the $2\times 2$ truncated CHF Hamiltonian.

Although a set of s.p. eigenstates $\{\epsilon_k^{(n)} (q_0-\Delta q),  \varphi_k^{(n)}(q_0-\Delta q)\}$ is numerically obtained by diagonalizing the full CHF Hamiltonian $h^{(n)}(q_0-\Delta q)$, a characteristic feature of interacting orbits at the $n$th iteration may be understood by the truncated Hamiltonian and $2\times 2$ unitary matrix $U^{(n)}$
\bea
\left( \begin{array}{c} \varphi_A ^{(n)} \\  \varphi_{A+1} ^{(n)} \end{array} \right) &=& \left( \begin{array}{cc} a^{(n)} & b^{(n)}  \\ d^{(n)} & c^{(n)}  \end{array} \right) \left( \begin{array}{c} \varphi_A(q_0) \\  \varphi_{A+1}(q_0) \end{array} \right)  \nonumber \\
&=&U^{(n)} \left( \begin{array}{c} \varphi_A(q_0) \\  \varphi_{A+1}(q_0) \end{array} \right).
\label{trans1}
\eea
In each diagonalization, $A$ and $A+1$ are always used to label the s.p. states in an energy increasing order.
To show a decisive effect of relative phase $a^{(n)}b^{(n)}$ on the non-convergence of the CHF calculation, we introduce four inter-dependent parameters in $U^{(n)}$ rather than a single independent parameter. The mixing parameters $b^{(n)}$ and $d^{(n)}$ measure a degree of interaction between two specific orbits.
Here and hereafter,  $\varphi_A ^{(n)}(q_0-\Delta q)$ and $h^{(n)}(q_0-\Delta q)$ at the $n$th iteration are simply expressed as $\varphi_A ^{(n)}$ and $h^{(n)}$.

\begin{figure}
\epsfxsize=7.1cm
\centerline{\epsffile{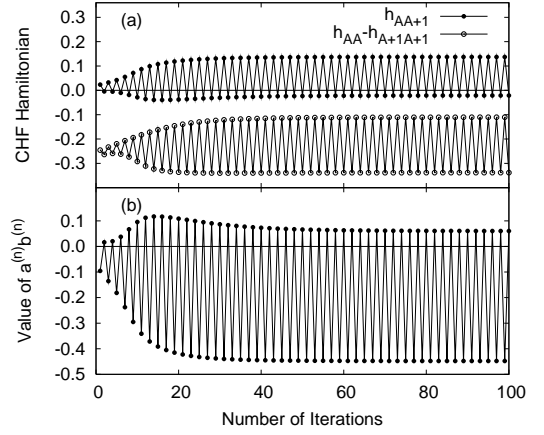}}
\caption{ \label{Hfh} For a given non-convergent case the off-diagonal Hamiltonian and difference of diagonal Hamiltonians (a) and
the value of $a^{(n)} b^{(n)}$ (b) as a function of iterative number.}
\end{figure}

In the $q_0$-representation, one gets the relation between two successive CHF Hamiltonians as
\bea
h_{A,A+1}^{(n+1)} &=& h_{A,A+1}^{(n)}+\{{a^{(n)}}{b^{(n)}}-{a^{(n-1)}}{b^{(n-1)}}\}\bar v_{A+1AAA+1}  \nonumber \\
&-&\{\Delta\lambda^{(n)}-\Delta\lambda^{(n-1)}\} Q_{A,A+1} ,\nonumber \\
h_{A,A}^{(n+1)}&=&h_{A,A}^{(n)}-\{{b^{(n)}}^2-{b^{(n-1)}}^2\}\bar v_{A+1AAA+1}   \nonumber \\
&-&\{\Delta\lambda^{(n)}-\Delta\lambda^{(n-1)}\} Q_{A,A} ,\nonumber \\
h_{A+1,A+1}^{(n+1)} &=& h_{A+1,A+1}^{(n)}+\{{b^{(n)}}^2-{b^{(n-1)}}^2\}\bar v_{A+1AAA+1}  \nonumber \\
&-&\{ \Delta\lambda^{(n)}- \Delta\lambda^{(n-1)} \}Q_{A+1,A+1},
\label{ite2}
\eea
where
\beq
\Delta\lambda^{(n)}=-2w{a^{(n)}}{b^{(n)}} Q_{A,A+1}-w{b^{(n)}}^2\{ Q_{A+1,A+1}-Q_{A,A}\},
\label{dlam}
\eeq
and the anti-symmetried two-body interaction $\bar v_{A+1AAA+1}$ is defined in the $q_0$-representation as
\bea
&&\bar v_{A+1AAA+1} \nonumber \\
&&=\sum_{\alpha\beta\gamma\delta}\varphi_{A+1\alpha}^{\dagger}(q_0)\varphi_{A\gamma}^{\dagger}(q_0) \bar v_{\alpha\gamma\beta\delta} \varphi_{\beta A}(q_0)\varphi_{\delta A+1}(q_0).\nonumber \\
\eea
Similarly, the $(n+2)$th off-diagonal CHF Hamiltonian is related with the $n$th through
\bea
&&h_{A,A+1}^{(n+2)}=h_{A,A+1}^{(n)}+\{{a^{(n+1)}}{b^{(n+1)}}-{a^{(n-1)}}{b^{(n-1)}}\}\nonumber \\
&& \bar v_{A+1AAA+1}-\{\Delta\lambda^{(n+1)}-\Delta\lambda^{(n-1)}\} Q_{A,A+1}.
\label{ite3}
\eea

Numerical values of $ h^{(n)}_{A,A} - h^{(n)}_{A+1,A+1}$ and $ h^{(n)}_{A,A+1}$ are shown in Fig.~\ref{Hfh}(a) for the non-convergent case with $\mu=150$ $\mbox{fm}^2$. Here the matrix elements are calculated in the $q_0$-representation at the edge point of convergence $q_0=178$ $\mbox{fm}^2$.
From this figure, one may observe that both components exhibit a staggering property around some averaged values.
It should be noticed that the off-diagonal hamiltonian $h^{(n)}_{A,A+1}$ changes its sign from iteration to iteration.
According to the perturbation theory for $2\times 2$ Hamiltonian, a sign of $a^{(n)} b^{(n)}$ of the lower state is given by that of $(h^{(n)}_{A,A+1})/(h^{(n)}_{A,A}-h^{(n)}_{A+1,A+1})$.
In Fig.~\ref{Hfh}(b), a numerical value of $a^{(n)} b^{(n)}$ is shown as a function of iterative number.
Comparing Fig.~\ref{Hfh}(a) with (b), one may recognize that the above discussion of perturbation theory is well justified by the numerical calculation.
Since the sign of $a^{(n)} b^{(n)}$ changes by iteration and iteration, one may deduce such a conclusion that the properties of the upper state $(-,-)_{A+1}$ and the lower state $(-,-)_A$ are inter-changed from one iteration to the next.

As seen from Eq.~{(\ref{ite2})}, $h_{A,A+1}^{(n+1)}$ becomes large because of the different signs between $a^{(n)}b^{(n)}$ and $a^{(n-1)}b^{(n-1)}$, whereas $h_{A,A+1}^{(n+2)}$ becomes small at the $(n+2)$th iteration because of the same sign of $a^{(n+1)}b^{(n+1)}$ and $a^{(n-1)}b^{(n-1)}$ in Eq.~{(\ref{ite3})}.
Since the quantum fluctuations coming from two-body residual interaction and quadrupole deformation become small in one iteration and large in the next, there appears the staggering property and they could not be approximated successfully by the one-body mean-field potential.
Physically one could understand the above situation as follows: two mean-fields, one characterized with $\varphi_A(q_0)$ and the other with $\varphi_{A+1}(q_0)$, interact too strongly by the two-body residual interaction to be approximated by a single mean-field.
Although the above phenomenon in Fig.~\ref{Hfh} has been known as ``ping-pong''~\cite{Dob00}, it should be noted that the present analytic understanding is given for the first time.

\begin{figure}
\epsfxsize=6.8cm
\centerline{\epsffile{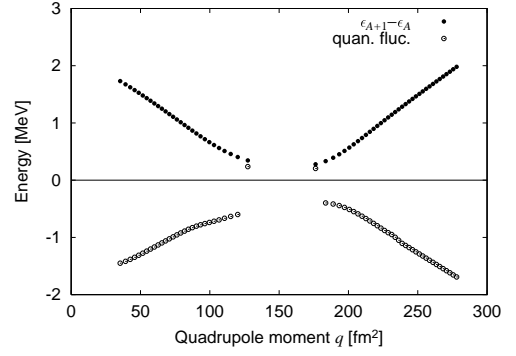}}
\caption{\label{Quan} The difference of s.p. energies and quantum fluctuation.}
\end{figure}

The above analytic and numerical results clearly indicate that the sign of $a^{(n)}b^{(n)}$ between two successive iterations is important whether or not the self-consistent CHF theory would be applicable.
Our next task is to make clear an analytic condition, which could tell us an existence of the mean-field.
The preceding discussion clearly indicates that the concept of CHF mean-field is applied successfully, 
provided there holds a condition
\beq
\frac{{h_{A,A+1}^{(n+1)}}}{{h_{A,A+1}^{(n)}}} \geq 0,
\label{cond1}
\eeq
whose most important part is evaluated for a case with $n=1$.
With the aid of Eqs.(\ref{ite2}) and (\ref{ite3}), it is easily shown that Eq.~(\ref{cond1}) is satisfied for any $n$ provided there holds Eq.~(\ref{cond1}) with $n=1$.
After some calculation, one may derive a relation
\bea
\label{cond}
&&\mbox{sign}\left\{ \frac{h_{A,A+1}^{(2)}}{h_{A,A+1}^{(1)}}\right\} =\mbox{sign}\{\epsilon_{A+1}(q_0)-\epsilon_{A}(q_0)-\bar v_{A+1AAA+1} \nonumber \\
&&-2w Q_{A,A+1}^2-w\Delta\mu (Q_{A,A}-Q_{A+1,A+1})-O(b) \}.
\eea
Here $O(b)$ contains a small mixing parameter $b^{(1)}$, and is expressed as
\bea
&&O(b)=-2{b^{(1)}}^{2}\bigl(\bar v_{A+1AAA+1}+2wQ_{A,A+1}^2\bigl)+ \nonumber \\
&& \frac{wb^{(1)}}{a^{(1)}}\bigl(1-2{b^{(1)}}^2\bigl)Q_{A,A+1}\bigl(Q_{A+1,A+1}-Q_{A,A}\bigl).
\eea
Condition (\ref{cond1}) for $n=1$ is then expressed as
\bea
&&\epsilon_{A+1}(q_0) - \epsilon_{A}(q_0)  \geq  \bar v_{A+1AAA+1}+2w{Q^2_{A,A+1}}\nonumber \\
&&+ w\Delta\mu (Q_{A,A}-Q_{A+1,A+1})+O(b).
\label{cond2}
\eea
Here the r.h.s. of condition (\ref{cond2}) with neglecting the higher-order effect $O(b)$ is the quantum fluctuation expressed as
$ \bar v_{A+1AAA+1}+2w{Q^2_{A,A+1}}+ w\Delta\mu (Q_{A,A}-Q_{A+1,A+1})$. The physical meaning of condition (\ref{cond2}) is clear: the two-body correlation between nucleons can be successfully incorporated into the mean field when the 
energy difference between two interacting orbits is not smaller than the quantum fluctuation coming from two-body residual interaction and quadrupole deformation. Similar derivation gives the opposite condition 
\bea
&&\epsilon_{A+1}(q_0) - \epsilon_{A}(q_0)  <  \bar v_{A+1AAA+1}+2w{Q^2_{A,A+1}}\nonumber \\
&&+ w\Delta\mu (Q_{A,A}-Q_{A+1,A+1})+O(b),
\label{cond3}
\eea
which states the breakdown of mean field.

\begin{figure}
\epsfxsize=6.8cm
\centerline{\epsffile{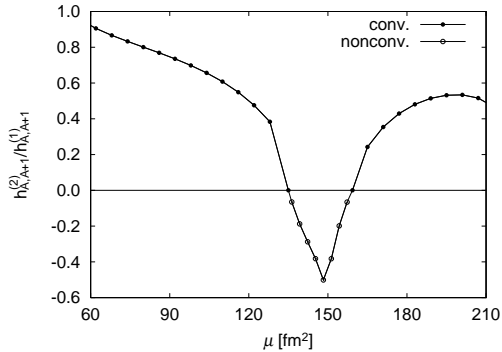}}
\caption{\label{Cond} The ratio of off-diagonal Hamiltonians between the first and second iterations for each given $\mu$.}
\end{figure}

Figure~{\ref{Quan}} shows the l.h.s of condition (\ref{cond2}) as well as the r.h.s. with neglecting the higher-order effect $O(b)$.
It is clear that our numerical results show the feasibility of the analytic condition~(\ref{cond2}).
Namely, both sides of condition~(\ref{cond2}) take almost the same value, when $q$ reaches to the edge points of the convergence.
Figure~{\ref{Cond}} depicts the ratio $h^{(2)}_{A,A+1}/h^{(1)}_{A,A+1}$, the l.h.s of Eq.~(\ref{cond1}) with $n=1$, as a function of quadrupole parameter $\mu$. One may see that the ratio is kept positive in convergent region, and reaches zero near the edge of it.
When one slightly decreases $\mu$ from the edge point, the ratio becomes negative where the convergence of CHF calculation is not achieved and condition~(\ref{cond3}) is satisfied.

The present work does not include the pairing correlation. Including the pairing correlation might make the situation so far discussed more complicated because of many dynamical competitions that exist not only between the $ph$-type two-body residual interaction and the HF potential, but also the $pp$-type two-body residual interaction and the pairing potential; their cross effects also must be considered. Moreover, the two mean fields characterized by different configurations are mixed up by the pairing correlation, and the $uv$ factor introduced by the BCS theory obscures the concept of the configuration.  Some numerical evidence of the above non-convergent feature in CHF-Bogoliubov (CHFB) theory has been discussed in Ref.~\cite{Guo2}
and its analytic derivation on an applicability of the CHFB theory near level crossing region is on progress.

Summarizing the present Letter, we have shown that the concept of CHF mean-field  breaks down near the level-repulsive region when the condition (\ref{cond2}) is not satisfied. Deriving the analytic condition, we make it clear that  the competition between one-body potential and quantum fluctuations mainly coming from two-body residual interaction plays an important role whether the self-consistent CHF mean field is realized or not.
Since the above result is obtained by using the constrained operator $\hat Q$ put by hand, further studies are needed whether an introduction of ``diabatic'' orbits \cite{Stru77, Dob00} really remedies the present difficulty or not. That is, the system does not like to be elongated nor contracted along a given direction of quadrupole deformation any more, but it likes to develop toward a direction chosen by itself.
Further microscopic investigation is needed to answer the very interesting conclusion in Ref.~\cite{RB89} by introducing {\it dynamical} constrained operators based on the self-consistent collective coordinate (SCC) method \cite{SCC1, SCC2}, because a diabolic point related to the level crossing indicates an existence of missing degree of freedom.

L. Guo acknowledges the financial support from the Alexander von Humboldt Foundation. E. G. Zhao acknowledges the support by NSFC under grant nos. 10375001 and 10435010.

\bibliography{se}
\end{document}